# Single-Mode Laser Arrays with Size-Invariant Loss Discrimination

Wenwen Liu[1, 2], Shuang Zhang[1,2,3,4.5]*

[1]New Cornerstone Science Laboratory, Department of Physics, The University of Hong Kong, Hong Kong

[2]State Key Laboratory of Optical Quantum Materials, The University of Hong Kong, 999077 Hong Kong, China

[3]Department of Electrical and Electronic Engineering, The University of Hong Kong, 999077 Hong Kong, China

[4]Quantum Science Center of Guangdong-Hong Kong-Macao Great Bay Area, 3 Binlang Road, Shenzhen, China

[5]Materials Innovation Institute for Life Sciences and Energy (MILES), HKU-SIRI, Shenzhen, People's Republic of China

*Corresponding authors. Email: shuzhang@hku.hk;

**ABSTRACT**. In conventional coupled oscillator lattices and laser arrays, expanding the active volume inevitably causes spectral crowding and degrades spatial uniformity, posing a long-standing barrier to high-power single-mode operation. We uncover a dissipative selection mechanism in sparse star networks that decouples modal selectivity from lattice size constraints. By engineering dissipative auxiliary couplings between a central hub and *N* peripheral resonators, the system establishes an exact star-Laplacian symmetry. This enforces a target eigenstate that is immune to all lossy channels while maintaining equal amplitude across all main nodes. All competing eigenstates break this symmetry, driving the auxiliary loss and opening a persistent, size-independent modal loss gap that prevents multi-mode competition. Realistic microring simulations validate the robustness of this dark-state isolation under electromagnetic dispersion and mechanism extension to arbitrary arrays is allowed when meets the system's exceptional point. By exploiting graph topology to bypass multi-mode competition, this work establishes a general route toward scalable high-power single-supermode laser arrays and robust state selection in coupled non-Hermitian systems.

## I. INTRODUCTION

Scaling the emission power and active aperture of semiconductor lasers while preserving single-mode operation is a long-standing challenge in coherent photonics [1-10]. In broad-area emitters and conventional coupled cavity lattices, increasing the physical footprint leads to spectral congestion, where the modal discrimination gap diminishes as the array size grows [10,11]. Consequently, dense mode competition and spatial hole burning inevitably trigger multi-mode filamentation and beam degradation at elevated pump levels [1–3]. Realizing a truly scalable high-power laser array requires two simultaneous conditions that are inherently difficult to reconcile: robust, persistent discrimination between the target state and its nearest competing supermodes, alongside a spatially uniform field profile across the entire lattice to maximize energy extraction without localized gain saturation and heat accumulation.

Recent advances in non-Hermitian photonics have provided powerful strategies for supermode engineering [11–20]. By engineering complex coupling matrices or introducing selective dissipation, unwanted supermodes can be preferentially suppressed through increased non-Hermitian decay rates. Notably, supersymmetric (SUSY) laser arrays utilize lossy partner networks to phase-match and damp higher-order transverse modes, isolating a fundamental lasing mode [21–30]. Concurrently, topological lattices, flat-band arrays, and non-Hermitian zero-mode lasers have harnessed edge states, sublattice symmetries, or protected bandgap spectra to achieve provide mode isolation [31–52]. Despite their elegance, existing architecture encounter significant scalability constraints and simultaneously achieving scalable mode discrimination and a spatially uniform supermode remains challenging. Specifically, many of the modes such as selected topological and defect modes are often spatially localized and exhibit nonuniform profiles [31–36,41,42]. Moreover, increasing the system size does not preserve the modal gap or ensure equal participation of all active elements [11,26,28,40].

Here, we propose a graph-theoretic approach to scalable single-supermode operation by exploiting the dissipative symmetries of a sparse star network. Starting from an ideal star topology where $N$ identical peripheral nodes couple to a central hub via fixed imaginary couplings, collective constructive interference singles out a low-loss target state, opening an imaginary modal gap that broadens in a square root scale with $N$. We physically implement these links by inserting lossy auxiliary resonators between the hub and peripheral cavities. In this architecture, the target supermode undergoes complete destructive interference within the lossy auxiliary cavities, rendering it a spatially uniform state immune to dissipation and granting it the lowest lasing threshold. Conversely, all competing supermodes violate this cancellation condition, drive the lossy auxiliaries, and experience strong suppression that preserves a persistent, size-independent modal discrimination gap. Full-wave simulations of microring resonator arrays validate the existence of a low-loss target state with a uniform spatial profile. The proposed dissipative star architecture thus provides a scalable framework for high-power coherent laser arrays and robust mode selection in distributed non-Hermitian systems.

## II. Ideal sparse selector

The central-cavity selection mechanism can be understood from an ideal star graph $S_{N+1}$ [Fig. 1(a)], comprising $N$ identical, mutually uncoupled peripheral cavities $|j\rangle$ $(j = 1, \dots, N)$ connected to a common central cavity $|C\rangle$ through equal, reciprocal imaginary couplings $i\kappa_c$. The state vector is expressed as $|\psi\rangle = c|C\rangle + \Sigma_j a_j\ |j\rangle$, and the system Hamiltonian takes the form:

$$H_{star} = \omega_0 I_{N+1} + i\kappa_c A_s \tag{1}$$

where $\omega_0$ is the uncoupled resonance frequency and adjacency matrix $A_s = \sum_{j=1}^{N}(\ |C\rangle\langle j|\ +\ |j\rangle\langle C|\ )$ specifies coupling topology. As illustrated in Fig. 1(b), this network supports three distinct mode families: a selected in-phase target supermode $\psi_+$with net gain, an $(N-1)$-fold degenerate manifold

of spatially non-uniform modes with zero imaginary energy, and a high loss radial-antiphase mode $\psi_-$. The target mode $\psi_+$ benefits from the lowest lasing threshold, and its separation from the nearest competing modes yields a normalized imaginary discrimination gap $\Delta_{\text{star}}/\kappa_c = \sqrt{N}$, establishing a scalable modal gap that broadens with array size (Fig. 1(c)).

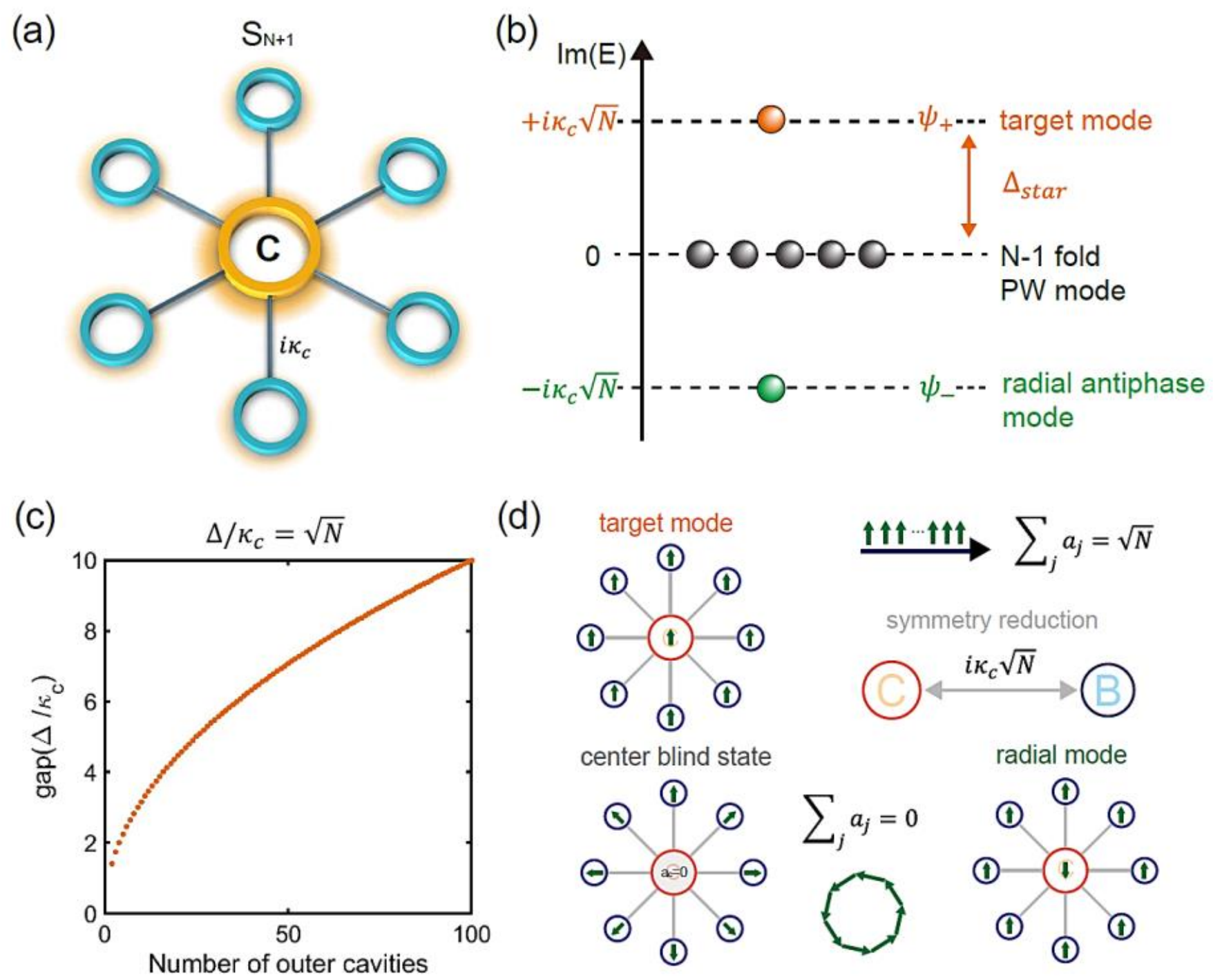

**Figure 1. Ideal fixed-link central-cavity selector**. (a) The star structure with one central cavity and $N$ outer rings. There are connected through one fixed reciprocal coupling $i\kappa_c$ .(b) The imaginary spectrum consists of the in-phase target $\psi_+$ at $+\kappa_c\sqrt{N}$, an $(N-1)$-fold PW manifold at zero, and the radial antiphase mode $\psi_-$ at $-\kappa_c\sqrt{N}$. The largest Im(E) has the lowest threshold. (c) The normalized gap grows as $\sqrt{N}$.(d) Eigenvector distribution of the three families of mode in the star network. For the in-phase target one , uniformly phased peripheral fields form the normalized collective state $|B\rangle$, which couples to the center $|C\rangle$ with the enhanced strength $i\kappa_c\sqrt{N}$. Peripheral configurations satisfying $\sum_j\langle j|q\rangle = 0$ cancel at the center and span an $(N-1)$ −dimensional center-blind manifold. The radial supermode remains center-visible with $\pi$ phase shift between the center and the outer cavities.

The selection mechanism becomes transparent when organizing the peripheral subspace by the cyclic permutation symmetry of the $N$ identical peripheral sites. Under a discrete rotation $R$ by $2\pi/N$, the normalized symmetry-adapted peripheral eigenvectors are

$$|q\rangle = \frac{1}{\sqrt{N}} \sum_{j=0,\dots,N-1} \exp\left(\frac{i2\pi qj}{N}\right) |j\rangle \tag{2}$$

with $R|q\rangle = \exp\left(\frac{-i2\pi q}{N}\right)|q\rangle$.The trivial representation ($q = 0$) corresponds to the uniform, peripheral in-phase state $|B\rangle = \frac{1}{\sqrt{N}} \Sigma_j |j\rangle$, whose fields interfere constructively at the central cavity. By contrast, each nontrivial sector ($q \neq 0$) exhibits an azimuthal phase winding that satisfies the local cancellation condition $\sum_j a_j = 0$. Because every peripheral resonator couples to the hub through an identical matrix element $i\kappa_c$ , the central coupling to a peripheral state is $\langle C|H_{star}|q\rangle = i\kappa_c / \sqrt{N}\sum_{j=0,\dots,N-1}\exp\left(\frac{i2\pi qj}{N}\right)|j\rangle = i\kappa_c\sqrt{N}\delta q_0$ . The central hub thus functions as a symmetry filter: all

phase winding (PW) modes ($q \neq 0$) decoupled from it, forming an $(N-1)$-dimensional center-blind subspace, whereas the uniform $q = 0$ state interacts exclusively to the hub with a collectively enhanced strength $i\sqrt{N}\kappa_c$. Consequently, the full $(N+1)$-dimensional Hamiltonian decomposes into block-diagonal symmetry sectors. Each nontrivial $q \neq 0$ sector forms an isolated one-dimensional block with zero central amplitude, spanning the $(N-1)$-fold degenerate manifold at energy $E_{q\neq 0} = \omega_0$. The only $2 \times 2$ block resides in the rotationally symmetric subspace spanned by $\{|C\rangle, |B\rangle\}$:

$$H_{q=0} = \omega_0 I + i\kappa_c \begin{bmatrix} 0 & \sqrt{N} \\ \sqrt{N} & 0 \end{bmatrix}$$

The corresponding eigenenergy is $E = \pm i\kappa_c\sqrt{N}$ with vectors $|\psi_\pm\rangle = (|C\rangle \pm |B\rangle)/\sqrt{2} = 1/\sqrt{2N}\left[\sqrt{N}, \pm 1, \pm 1, \dots, \pm 1\right]^T$ (Fig. 1(d)). The target state $\psi_+$ is uniform in phase both azimuthally and radially, benefiting from maximum collective constructive interference that lifts its imaginary energy to $+\kappa_c\sqrt{N}$ (lowest lasing threshold). A representative PW state winds in phase around the outer ring and has no central field. Modal discrimination thus arises through two distinct interference mechanisms: the PW ($q \neq 0$) center-dark modes transparent to the hub due to azimuthal destructive interference ($\sum_j a_j = 0$), while the radial antiphase mode $\psi_-$ couples strongly to the hub but acquires maximum dissipation ($-\kappa_c\sqrt{N}$) owing to a $\pi$ phase shift between the center and the peripheral array.

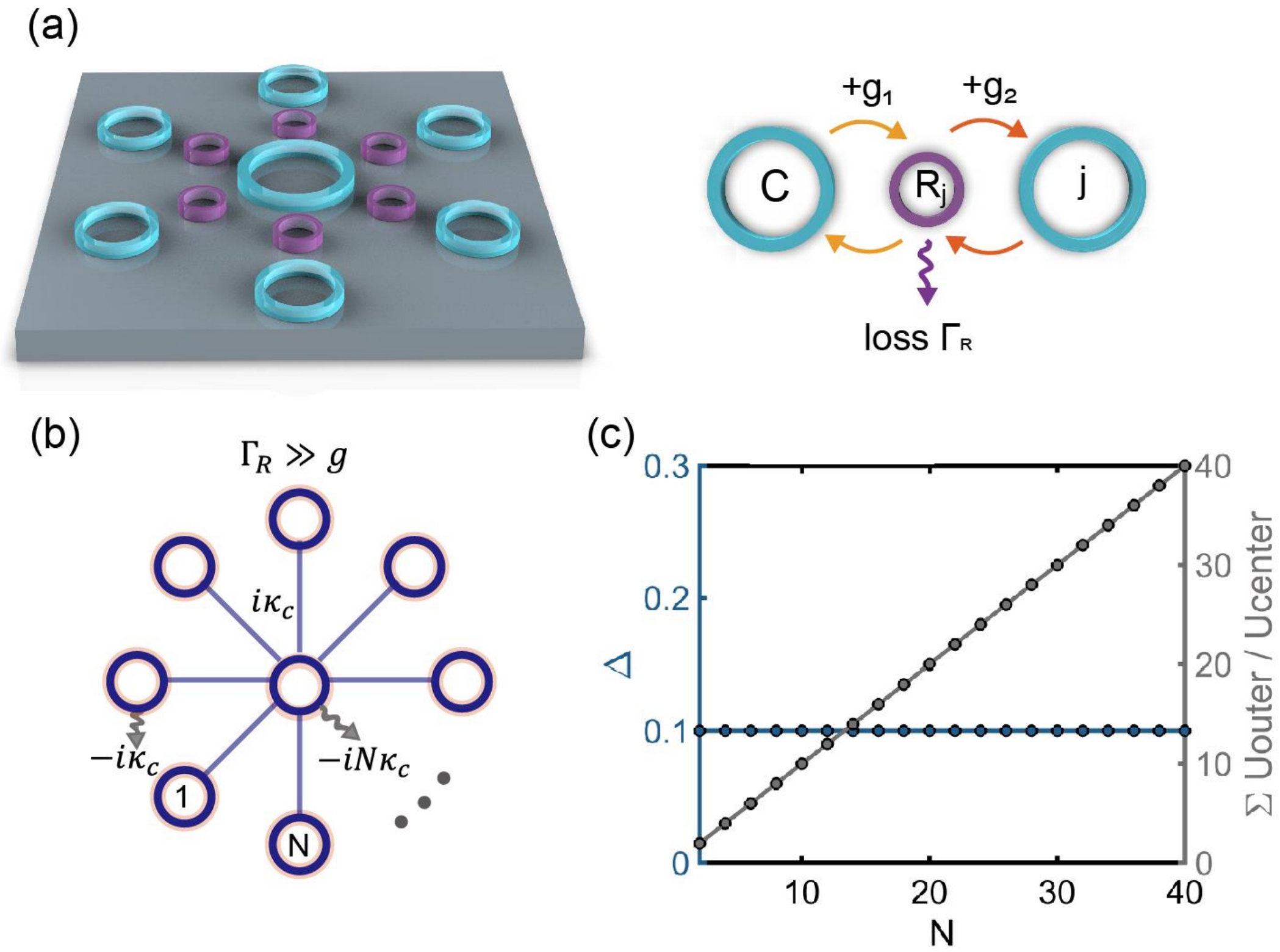

**Figure 2**. **Effective passive lossy-auxiliary implementation**. (a) (left) Full star structure with $N$ lossy auxiliary resonator $R_j$ (purple) inserted between the central cavity and the outer ring (blue). (right) Zoom in structure of one link. (b) Effective model with condition $\Gamma_R \gg g$. The center receives $-iN\kappa_c$ loss effectively while the outers receive $-i\kappa_c$ loss. All the cavities hold the same amplitude at the target mode. (c) Target-field distribution efficiency (grey) and gap (blue) with the increase of system size for the effective model at fixed $\Gamma_R$ and $g$.

## III. Star-Laplacian implementation via lossy auxiliaries

To physically realize the reciprocal imaginary hoppings postulated in the ideal star model of Eq. (1) without elaborate active gain engineering, we introduce $N$ lossy auxiliary resonators $R_j$ $(j = 1, \ldots, N)$, each mediating the interaction between the central hub $C$ and peripheral cavity $j$ [Fig. 2(a)]. The total $(2N+1)$-dimensional Hilbert space partitions naturally into the main-cavity subspace $M = \{|C\rangle, |1\rangle, \ldots, |N\rangle\}$ and the auxiliary reservoir subspace $A = \{|R_1\rangle, \ \ldots\ , |R_N\rangle\}$. In a local mode basis where the coherent couplings from each auxiliary resonator to the hub $(g_1)$ and its corresponding peripheral cavity $(g_2)$ are real, the full Hamiltonian on resonance takes the exact block form:

$$H_{\text{full}} = \begin{pmatrix} H_{MM} & H_{MA} \\ H_{AM} & H_{AA} \end{pmatrix} = \begin{pmatrix} 0_{N+1} & Q \\ Q^T & -i\Gamma_R I_N \end{pmatrix} \tag{3}$$

where $\Gamma_R$ is the engineered auxiliary decay rate and $Q = \begin{pmatrix} g_1 1_N^T \\ g_2 I_N \end{pmatrix}$ is the $(N+1) \times N$ coupling matrix. For an arbitrary main-cavity field vector $a = (a_c, a_1, \ldots, a_N)^T$, the drive on auxiliary resonator $R_j$ is $[Q^T \boldsymbol{a}]_j = g_1 a_C + g_2 a_j$. Choosing opposite coupling signs, $g_1 g_2 < 0$, introduces a relative $\pi$ phase difference. Destructive interference then eliminates auxiliary excitation when $a_C/a_j = -g_2/g_1 > 0$, thereby selecting an in-phase supermode.

For balanced coupling amplitude $g_1 = -g_2 = g$, the interaction matrix simplifies to $Q = gP$ with each column of $P$ contains $+1$ at the hub and $-1$ at each peripheral node. In the fast-decay limit ($\Gamma_R \gg g$), eliminating the auxiliary subspace $A$ via Feshbach-Schur projection yields the low-energy effective Hamiltonian:

$$H_{eff} = H_{MM} - Q H_{AA}^{-1} Q^T = -i\kappa_c L_s \tag{4}$$

where $L_s$ is the star-graph Laplacian measuring the branch difference. Comparing with Eq. (1), it only adds the diagonal part with degree-dependent loss associated with connected cavity while maintaining (Fig. 2(b)). This diagonal loss precisely counterbalances the structural degree asymmetry of the network, preserving the zero-loss target supermode ($a_C = a_1 = \cdots = a_N$) in the null space of the Laplacian ($L_s \mathbf{a} = 0$) while converting the size-dependent scaling of the ideal star [Fig. 1(c)] into a persistent, size-independent modal discrimination gap [Fig. 2(c)].

Beyond the adiabatic approximation $\Gamma_R \gg g$, the full $2N+1$-dimensional Hamiltonian remains invariant under the simultaneous cyclic permutation of the identical peripheral and auxiliary resonators.: $[H_{\text{full}}, \mathcal{R}] = 0$. We therefore introduce the symmetry-adapted states $|P_q\rangle = \frac{1}{\sqrt{N}} \sum_{j=0}^{N-1} e^{i2\pi qj/N} |j\rangle$ and $|A_q\rangle = \frac{1}{\sqrt{N}} \sum_{j=0}^{N-1} e^{i2\pi qj/N} |R_j\rangle$. Since the phase sum vanishes for every nontrivial representation, the central cavity couples only to the uniform auxiliary state as $\langle C | H_{\text{full}} | A_q \rangle = \frac{g_1}{\sqrt{N}} \sum_{j=0}^{N-1} e^{i2\pi qj/N} = \sqrt{N} g_1 \delta_{q0}$. Thus, the central cavity belongs exclusively to the $q = 0$ sector. After the discrete Fourier transformation, the Hamiltonian separates exactly into one rotationally symmetric sector and $N-1$ phase-winding sectors:

$$H_{q=0} = \begin{pmatrix} 0 & 0 & \sqrt{N} g_1 \\ 0 & 0 & g_2 \\ \sqrt{N} g_1 & g_2 & -i\Gamma_R \end{pmatrix}, H_{q \neq 0} = \begin{pmatrix} 0 & g_2 \\ g_2 & -i\Gamma_R \end{pmatrix}. \tag{5}$$

The $q = 0$ sector contains both the target state and the radial family. To identify the target state, consider a general main-cavity state $c|C\rangle + p|P_0\rangle$ in this sector. It drives the uniform auxiliary state with amplitude $\sqrt{N}\, g_1 c + g_2 p$. Setting this amplitude to zero gives the normalized dark state and can have $|D_0\rangle = \frac{-g_2|C\rangle+\sqrt{N}\, g_1|P_0\,\rangle}{\sqrt{g_2^2+Ng_1^2}}$ with $E_0 = 0$. This state satisfies the local cancellation condition $g_1 a_C + g_2 a_j = 0$ on every branch and therefore has exactly zero auxiliary amplitude for arbitrary $\Gamma_R$. For balanced couplings $g_1 = -g_2$, it reduces to the uniform in-phase state over all main cavities. More generally, the ratio $|g_1/g_2|$ redistributes the target-mode energy between the peripheral array and the hub according to

$$\frac{U_{\mathrm{outer}}}{U_{\mathrm{center}}} = N\left(\frac{g_1}{g_2}\right)^2 \qquad (6)$$

without changing the nearest competing loss when $g_2$ and $\Gamma_R$ are fixed.

The other state in the $q = 0$ main-cavity subspace is the bright state $|B_0\rangle = \frac{\sqrt{N}\, g_1|C\rangle+g_2|P_0\,\rangle}{\sigma_R}$ which is orthogonal to $|D_0\rangle$. It couples to $|A_0\rangle$ with strength $\sigma_R = \sqrt{Ng_1^2 + g_2^2}$, forming the radial lossy dimer with eigenvalues $E_{R,\pm} = \frac{-i\Gamma_R \pm \sqrt{4\sigma_R^2 - \Gamma_R^2}}{2}$.

For every $q \neq 0$, the phase sum vanishes, so $|A_q\rangle$ has no coupling to the hub. The pair $\{|P_q\rangle, |A_q\rangle\}$ therefore forms a closed peripheral-auxiliary dimer with exactly zero central-cavity amplitude. These modes are nevertheless locally bright because $a_C = 0$ gives the branch drive $g_2 a_j \neq 0$. All $N-1$ phase-winding sectors consequently have the same coupling $\sigma_D = |g_2|$ and eigenvalues $E_{D,\pm} = \frac{-i\Gamma_R \pm \sqrt{4g_2^2 - \Gamma_R^2}}{2}$. This phase-sector decomposition is equivalent to the singular-value spectrum $\sigma_0 = 0, \sigma_D = |g_2|, \sigma_R = \sqrt{Ng_1^2 + g_2^2}$, while directly exposing the underlying interference mechanism.

The spectrum consequently consists of three physical families: the exact auxiliary-dark target state, the radial $q = 0$ doublet, and the $(N-1)$-fold phase-winding doublets. Since $\sigma_R \geq \sigma_D$, the least-lossy competitors are always the phase-winding modes, and the exact modal loss gap is

$$\Delta = \frac{\Gamma_R - \sqrt{\Gamma_R{}^2 - 4g_2{}^2}}{2} \qquad (7)$$

which means the modal discrimination depends solely on the local peripheral link parameter, i.e., the gap $\Delta$ is completely independent of the emitter count *N*. Its dependence on $\Gamma_R$ and $g_2$ reveals an exceptional line at $g_2 = \Gamma_R/2$, beyond which the imaginary gap saturates (Fig. 3(c)). This behavior follows directly from the effective non-Hermitian dimer associated with each singular-value sector. An exceptional point (EP) occurs at $2\sigma = \Gamma_{\mathrm{R}}$, where the two eigenvalues and eigenvectors coalesce. The radial mode, with the collectively enhanced singular value $\sigma_{\mathrm{R}}$, reaches this condition first as the link strength increases (Fig. 3(d)). The PW sectors subsequently reach their EP at $\sigma_{\mathrm{D}} = \frac{\Gamma_{\mathrm{R}}}{2}$. Below this threshold, increasing $|g_2|$ drives the least-lossy competitor farther down the imaginary-energy axis, so $\Delta$ grows according to Eq. (9). Above it, the paired modes split in real frequency while retaining the decay rate $\frac{\Gamma_{\mathrm{R}}}{2}$. The PW EP therefore marks the transition from increasing loss discrimination to a saturated modal loss gap, beyond which stronger coupling provides no further improvement. Extension to arbitrary arrays of such mechanism would not change $\Delta$ when EP condition is met. As shown in Fig. 3 (e-f), the gap retains $\Delta = \frac{\Gamma_{\mathrm{R}}}{2}$ with fixed $g/\Gamma_R$=1. (see more details in Supplemental Material).

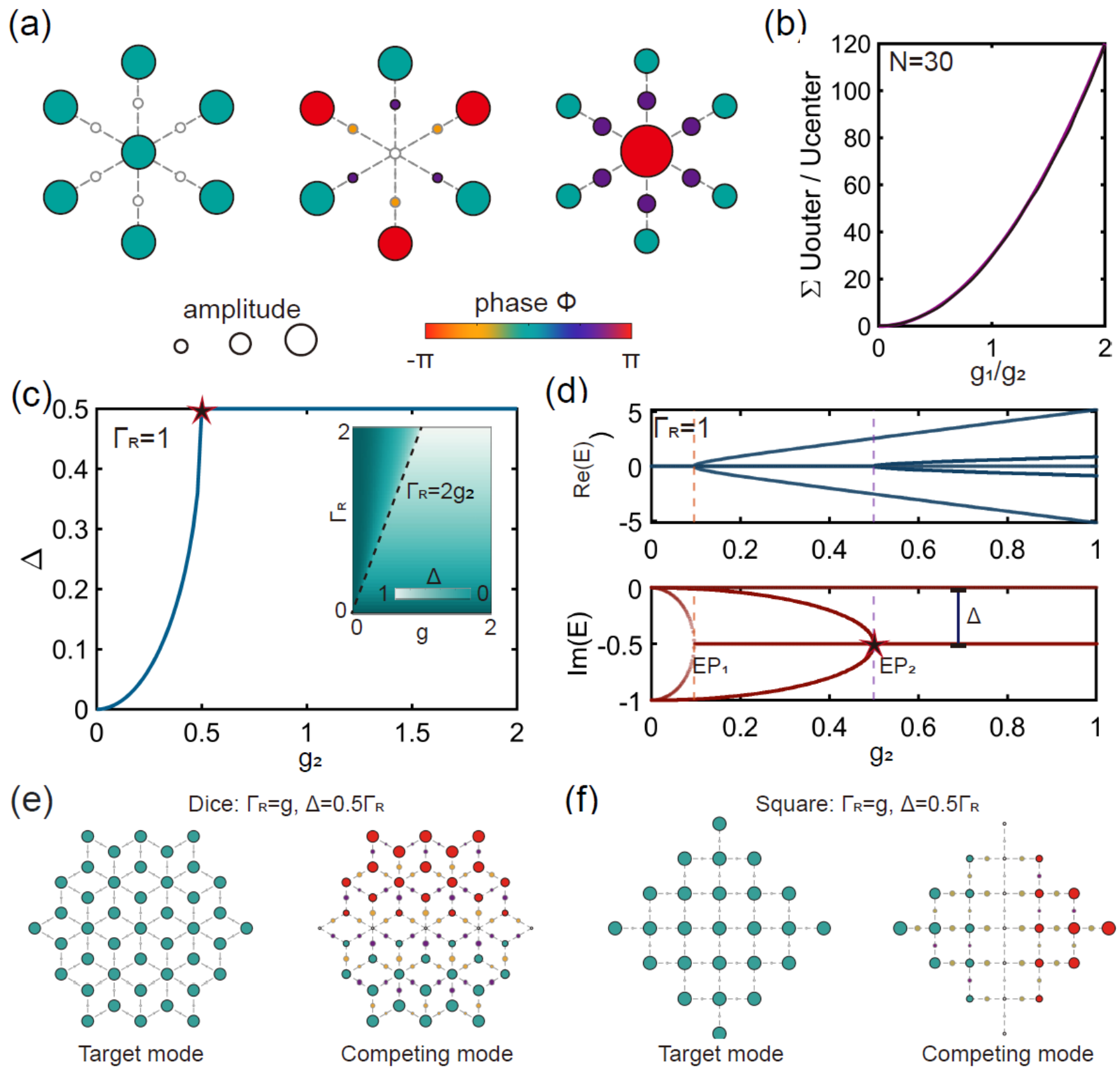

**Figure 3. Characteristic of the full star structure**. (a) Field distribution for the full 13-site network. The uniform Laplacian target has equal main-cavity amplitudes and exactly zero auxiliary field while PW competitors drive the auxiliaries. (b) The field located at center cavity would decrease with the increase of $\frac{g_1}{g_2}$. (c-d) The gap would increase before touching the EP line $\Gamma_R = 2g_2$ (c). With fixed $\Gamma_R = 1$, the real part and imaginary part of eigenenergy occurs EPs with the increase of $g_2$(d). (e-f) Mechanism extension considering the examples of dice lattice (e) and square lattice (f). The gap remains $\frac{\Gamma_R}{2}$ with fixed coupling coefficient.

## IV. Verification based on realistic design

To validate the theoretical framework, we investigate a microring resonator system operating near the telecommunication wavelengths (1587 nm), as shown in Fig. 4. The device consists of a central ring, $N$ identical peripheral rings, and $N$ lossy auxiliary racetrack resonators positioned as links. All resonators share a refractive index of 3 with an additional material loss of $\Gamma = 0.0026$ introduced specifically into the auxiliary cavities.

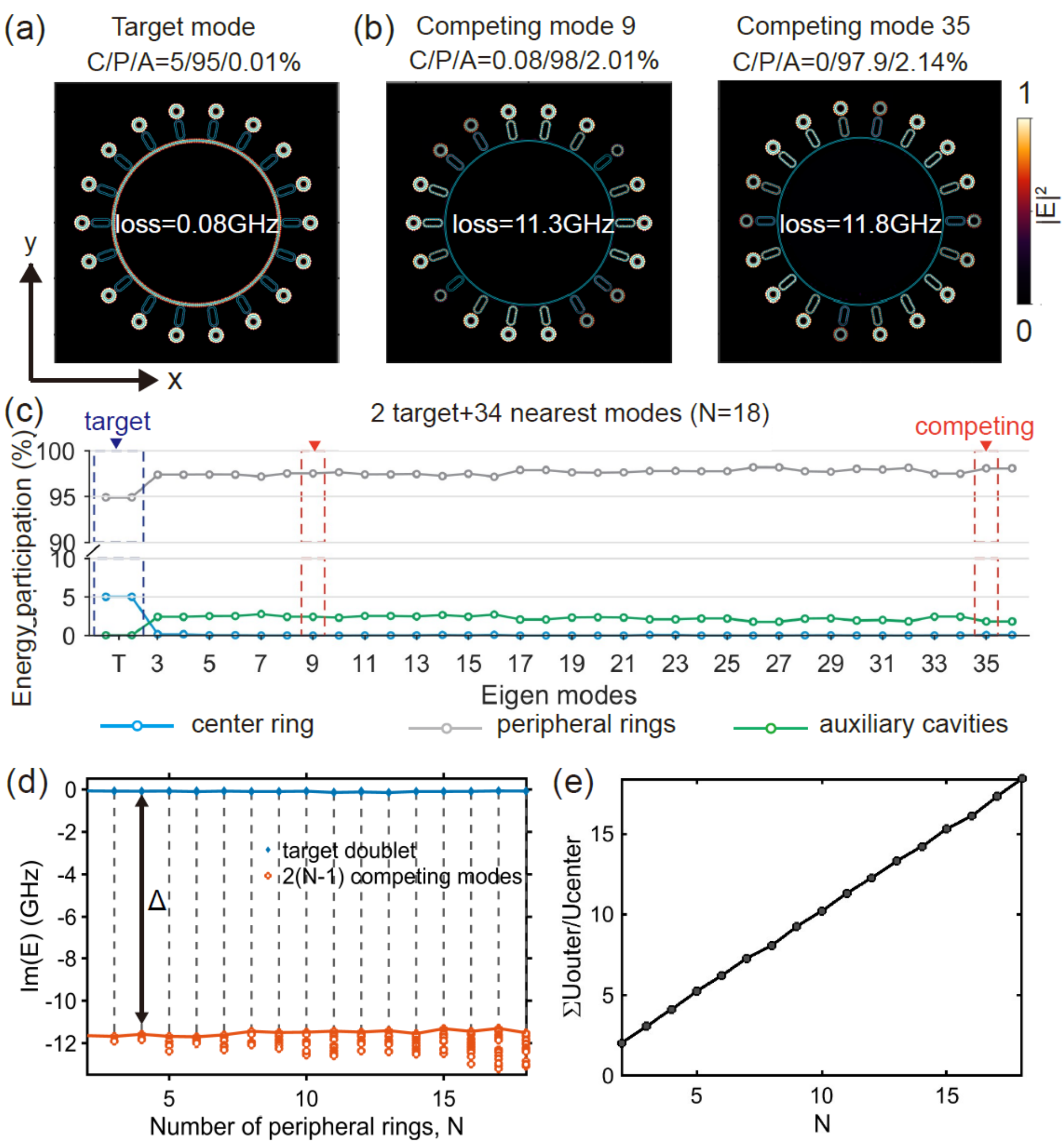

**Figure 4. Implementation in a microring resonator system**. (a-b) Normalized Energy distribution spectrum for the target mode (a) and two representative competing (b) modes. The competing modes are represented by standing-wave-like combinations of the nearly degenerate $q$ and $-q$ states. (c) Energy participation at different rings for all the target modes and competing modes (including all CW and CCW modes). Nearly no energy leak into the auxiliary cavities at the target modes while the competitors excite the auxiliaries and contribute to the imaginary gap. (d) The simulation result shows the imaginary gap remains almost unchanged with the increase of system size. (e) The ratio of energy at outer rings for the target mode grows with system size.

The central ring has a radius of $R = 1482nm$ with a width of $w_c = 200nm$, while the periphercal rings feature a radius of $r = 887nm$ and a width of $w_p = 300\mu m$. These dimensions are tuned to align their resonances, with the central and peripheral rings operating in modes $m = 126$ and $m = 9$, respectively. Similarly, the auxiliary racetrack are tuned to the same resonance. Full-wave electromagnetic field calculation directly resolves the complex eigenfrequency and field distribution for each supermode, allowing us to compute energy participation by integrating the field energy over the center ($C$), the complete peripheral array ($P$), and all auxiliary cavities ($A$).

For an array size of $N = 18$, the target states emerge as a degenerate doublet with clockwise (CW) and counterclockwise (CCW) circulation (Fig. 4(c)). These target modes exhibit an exceptionally low loss of 0.08 GHz, accompanied by an energy participation of $C/P/A = 5.0/94.9/0.01$ in percentage [Fig. 4(a)]. These values align closely with the theoretical uniform

Laplacian eigenvector predictions, which dictates a central participation fraction of $\frac{1}{N+1} = 5.26\%$ and a total peripheral fraction of $N/(N+1) = 94.74$. The negligible energy leaking into the auxiliary cavities confirms that the local destructive-interference condition remains robustly satisfied in the full-wave structure. In Fig. 4(c), Modes 3 to 38 correspond to the nearest competing states, with two representative field distributions shown in Fig 4(b). As the phase-winding states $q$ and $-q$ are degenerate, the full-wave eigensolver may return a standing-wave-like superposition $a_j = Ae^{iq\theta_j} + Be^{-iq\theta_j}$with intensity modulation. Importantly, both components have zero peripheral sum, so their superposition remains decoupled from the central cavity. These competing modes suffer a substantially higher loss of approximately 11.5 GHz. Their participation ratios $C/P/A = 0.08/98.1/2.01$ reveal the complementary selection mechanism: these modes are almost dark at the central ring but possess an azimuthally nonuniform peripheral field. The resulting field imbalance across the main-cavity links drives the auxiliary resonators, increasing auxiliary participation and subjecting the competing modes to additional dissipation.

As the system size $N$ increases, the extracted modal gap remains approximately constant over the simulated range [Fig. 4(d)], consistent with the size-independent gap predicted by the passive star-Laplacian model. Meanwhile, for the target modes, the ratio of total peripheral energy to central-ring energy increases linearly [Fig. 4(e)], as expected for equal energy participation in each main ring. The results therefore confirm the core device-level concept: the architecture successfully sustains a low-loss, auxiliary-dark target mode during array scaling, whereas nonuniform competitors incur finite auxiliary loading and are preferentially suppressed.

To conclude, we have proposed a star-Laplacian auxiliary-mediated mode selector for scalable single-supermode operation with a uniform spatial profile. The system supports a non-vanishing modal gap with growing system size. Its passive implementation produces an auxiliary-dark target state with equal participation in all main cavities, while competing modes experience additional loss. Implementation in microring resonator system confirms strong mode discrimination, a size-independent gap, and a peripheral-to-central energy ratio that increases linearly with system size. These findings establish a route toward scalable laser arrays with uniform single-supermode operation.